\documentclass[usenatbib]{mn2e}

\usepackage{epsfig}
\usepackage{mathrsfs,aas_macros}

\def\Lya{Ly$\alpha$~} 
\def\Lyb{Ly$\beta$~}

\def\HI{\hbox{H~$\rm \scriptstyle I\ $}}
\def\HII{\hbox{H~$\rm \scriptstyle II\ $}}

\title[A closer look at quasar near-zones]{A closer look at using quasar
near-zones as a probe of neutral hydrogen in the intergalactic medium}

\author[J.S. Bolton \& M.G. Haehnelt] {James S.
  Bolton$^{1}$ \& Martin G. Haehnelt$^{2}$ \\
  $^1$ Max Planck Institut f{\"u}r Astrophysik, Karl-Schwarzschild
  Str. 1, 85748 Garching, Germany \\
  $^2$ Institute of Astronomy, University of Cambridge, Madingley
  Road, Cambridge,  CB3 0HA \\}

\begin{document}

\date{09 July 2007}

\maketitle

\label{firstpage}

\begin{abstract}
We examine a large set of synthetic quasar spectra to realistically
assess the potential of using the relative sizes of highly ionized
near-zones in the \Lya and \Lyb forest  as a probe of the neutral
hydrogen content of the intergalactic  medium (IGM) at $z>6$.  The
scatter in the relative near-zone size  distribution, induced by
underlying fluctuations in the baryonic density field and the
filtering of ionizing radiation, is considerable even for fixed
assumptions about the IGM neutral fraction.  As a consequence, the
current observational data cannot distinguish between an IGM which is
significantly neutral or highly ionized just above $z=6$.  Under
standard assumptions for quasar ages and ionizing luminosities, a
future sample of several tens of high resolution \Lya and \Lyb
near-zone spectra should be capable of distinguishing between a volume
weighted neutral hydrogen fraction in the IGM which is greater or less
than 10 per cent.  

\end{abstract}
 

\section{Introduction}

Quasar absorption spectra are currently the premier observational
probe of the hydrogen reionization epoch.  The average amount of Lyman
series absorption observed in quasar spectra is consistent with an
intergalactic medium (IGM) with a volume weighted neutral
hydrogen fraction in excess of $\langle f_{\rm HI} \rangle_{\rm V}\sim
10^{-3.5}$ at $z\simeq 6$ (\citealt{Fan06t,Becker06b}).  However,
obtaining more stringent constraints on the IGM neutral hydrogen
fraction using this technique at $z>6$ is extremely difficult; larger
neutral fractions produce saturated Lyman series absorption
over substantial regions in the spectra of these quasars, the
observational manifestation of which is the \cite{GunnPeterson65} trough.

Consequently, several techniques which may be sensitive to larger
values of $\langle f_{\rm HI} \rangle_{\rm V}$ have been proposed (see
\citealt{Fan06rv} and references therein). The sizes of transparent
regions observed immediately blueward of the \Lya emission line of
$z>6$ quasars, if equivalent to the size of the \HII regions
surrounding the quasars, should be proportional to the ambient IGM
neutral hydrogen fraction when adopting assumptions for the quasar age
and ionizing luminosity (\citealt{CenHaiman00,MadauRees00,YuLu05}).
Under these assumptions, these regions are consistent with $\langle
f_{\rm HI} \rangle_{\rm V}>0.1$, just above $z=6$
(\citealt{WyitheLoeb04,Wyithe05}), although a recent independent
analysis including modelling of the Gunn-Peterson trough damping wing
by \cite{MesingerHaiman06} using many simulated quasar sight-lines
favours a slightly lower limit of $\langle f_{\rm HI} \rangle_{\rm
V}>0.033$.

However, recent numerical studies which correctly model the radiative
transfer of ionizing photons around these quasars indicate that
interpreting the sizes of these highly ionized \Lya near-zones with
respect to the neutral hydrogen content of the IGM is complicated
(\citealt{BoltonHaehnelt07,Maselli07,Lidz07}).  There are two main
reasons for this, in addition to the uncertain quasar age and ionizing
luminosity.  Firstly, for small neutral fractions the \Lya near-zone
resembles the classical proximity zone of a luminous quasar embedded
in a highly ionized IGM  (\citealt{Bajtlik88}) and does not correspond
to the extent of an \HII region expanding into a substantially neutral
IGM.  The observationally identified \Lya near-zone sizes can then
substantially underestimate the size of the region of enhanced
ionization surrounding the quasar, leading to an overestimate of the
IGM neutral hydrogen fraction.  Secondly, even for fixed assumptions
about the ionization state of the IGM, variations in the IGM density
and the ionizing background along different quasar sight-lines
combined with radiative transfer effects lead to a considerable
scatter in the near-zone sizes.  Drawing robust constraints on
$\langle f_{\rm HI} \rangle_{\rm V}$ from only a small sample of
spectra is therefore difficult.  \cite{BoltonHaehnelt07} found the
observed near-zone sizes at $z>6$ are consistent with neutral
hydrogen fractions as small as $\langle f_{\rm HI}  \rangle_{\rm V} =
10^{-3.5}$.

Nevertheless, further progress may be possible by also considering the
size of the corresponding \Lyb near-zone.  The edge of the \Lyb
near-zone should trace the position of the quasar \HII ionization
front (IF) to smaller volume averaged neutral hydrogen fractions in
the ambient IGM relative to the \Lya near-zone.  The difference in the
sizes of these regions thus holds additional information on the IGM
neutral hydrogen fraction
(\citealt{MesingerHaiman04,BoltonHaehnelt07}, hereafter BH07a).  In
principle, the ratio of \Lyb to \Lya near-zone sizes should increase
from unity to a maximum value of $\sim 2.5$ as the IGM neutral
hydrogen fraction decreases (BH07a).  In practise, however, the ratio
can be much smaller than $2.5$ even for very small neutral fractions
and it will exhibit a large scatter due to fluctuations in the IGM
density along the line of sight.

In this letter we extend the work of BH07a to examine the use of \Lya
and \Lyb near-zones as a probe of  $\langle f_{\rm HI} \rangle_{\rm
V}$.  Our work closely follows on from the concepts discussed in
BH07a, although the analysis presented here differs in three important
ways.  We analyse several hundred different simulations of the
radiative transfer of ionizing photons around $z>6$ quasars, greatly
improving the synthetic spectra statistics.  An approximate treatment
of an inhomogeneous ionizing background, expected towards the tail end
of hydrogen reionization (\citealt{Fan06t,Wyithe06}) is now also
included within the radiative transfer simulations.  Lastly we adopt a
more robust method for defining the sizes of both \Lya and \Lyb
near-zones.  As a consequence, we are able to more realistically
assess the potential of using quasar near-zones as a probe of the IGM
neutral hydrogen fraction beyond $z=6$.

\section{Simulations}
\subsection{Radiative transfer implementation and the IGM density distribution}

We use the one dimensional, multi-frequency photon-conserving
algorithm described and tested in BH07a to compute the radiative
transfer of ionizing photons around quasars.  This is combined with
density distributions drawn from the $400^{3}$ hydrodynamical
simulation of BH07a, run using the parallel {\tt TREE-SPH} code {\tt
GADGET-2} (\citealt{Springel05}).  The ten most massive haloes in the
simulation at $z=6.25$ were identified and baryon density
distributions were extracted in different orientations around them.
Continuous density distributions $150h^{-1}$ comoving Mpc in length
were then constructed using the halo density distributions combined
with other sight-lines drawn randomly from the simulation volume.  One
end of the density distributions always lies at the centre of one of the
identified haloes, where the quasar is assumed to reside during the
radiative transfer calculation.  This process was repeated to produce
240 unique sight-lines. 

Synthetic \Lya and \Lyb spectra are constructed from the output of the
radiative transfer simulations following standard procedure ({\it
e.g.} BH07a).  The raw synthetic spectra are processed to resemble
data obtained with the Keck telescope Echelle Spectrograph and Imager
(ESI).  The spectra are convolved with a Gaussian with a full width
half maximum (FWHM) of $66\rm~km~s^{-1}$ and rebinned onto pixels of
width $3.5\rm~\AA$ ($R\sim 2500$). Gaussian distributed noise is then
added with a total signal-to-noise ratio of $20$ per pixel at the
continuum level and a constant read out signal-to-noise of $80$ per
pixel.  {We also process some of the spectra to resemble data taken
with the Keck telescope High Resolution Echelle Spectrometer (HIRES).
These are convolved with a Gaussian with a FWHM of $7\rm~km~s^{-1}$ and
rebinned onto pixels of width $0.25\rm~\AA$ ($R\sim 35000$).  Noise is
then added with a total signal-to-noise ratio of $20$ per pixel at the
continuum level and a constant read out signal-to-noise of $100$ per
pixel.

Random sight-lines were also drawn from the simulation at $z=5.12$ to
model the foreground \Lya forest superimposed on the \Lyb absorption
at $z=6.25$.  These were spliced together to produce a $z=5.12$ \Lya
spectrum with the same wavelength coverage as the \Lyb spectrum at
$z=6.25$.  Lastly, the foreground \Lya optical depths are rescaled
({\it e.g.} \citealt{BoltonHaehnelt07b}) to reproduce the observed mean
flux of the \Lya forest at $z=5.12$, $\langle F \rangle=0.149$
(\citealt{Songaila04}).

\subsection{Initial conditions and the fluctuating ionizing background model}

We run a total of 480 radiative transfer simulations arranged into two
groups of 240 simulations.  The density distributions for both groups
have initial neutral hydrogen fractions distributed uniformly in the
range $-4 \la \log \langle f_{\rm HI} \rangle_{\rm V} \la 0$, set by
assuming the IGM is in ionization equilibrium with an ionizing
background with a power-law spectral index, $\alpha_{\rm b}=3$, below
the Lyman limit.  For the first group of 240 simulations the ionizing
background is assumed to be spatially uniform, as in BH07a.

\begin{figure}
\begin{center} 
  \includegraphics[width=0.4\textwidth]{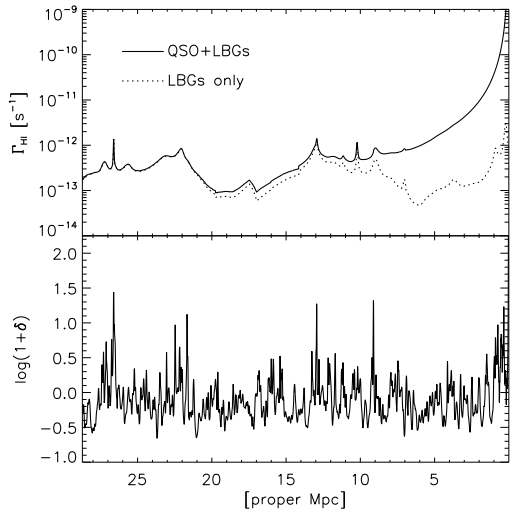}
  \caption{{\it Top:} Example of the spatially fluctuating hydrogen ionization
  rate along a simulated quasar sight-line.  The quasar is situated at
  the right hand side of the diagram.  The dotted line shows the
  ionizing background due to galaxies only and the solid line also
  includes the contribution from the quasar computed using the
  radiative transfer implementation. {\it Bottom:} The corresponding
  baryonic overdensity along the sight-line.  The ionizing
  background is enhanced in overdense regions due to the clustering of
  ionizing sources. }
\label{fig:radfield}
\end{center} 
\end{figure} 

However, a spatially uniform ionizing background is likely to be a
poor approximation at the tail end of  reionization and this
assumption may have an impact on the near-zone sizes derived from
simulations (\citealt{Lidz07,AlvarezAbel07}).  Therefore, for the
second group of simulations we construct spatially fluctuating
ionizing backgrounds using observationally determined Lyman break
galaxy (LBG) luminosity functions at high redshift. We do this using
the model of \cite{BoltonHaehnelt07b}, to which we refer the reader
for further details.  An example of the fluctuating photo-ionization
rate per hydrogen atom, $\Gamma_{\rm HI}$, along a single sight-line
at $z=6.25$ is shown in Figure~\ref{fig:radfield}.  The background
value of $\Gamma_{\rm HI}$ due to galaxies is several orders of
magnitude larger in overdense regions due to the clustering of
ionizing sources, although it is still dominated by the quasar
radiation field within $10$ proper Mpc of the host halo. In order to
reproduce different initial values of $\langle f_{\rm HI} \rangle_{\rm
V}$ in our simulations we rescale the mean of the fluctuating
ionization rate along each sight-line by the appropriate amount.
Note, however, that this simple model only applies during the
post-overlap phase of reionization  ({\it cf.} \citealt{Gnedin00})
when the ionizing photon mean free path is larger than the typical
separation between ionizing sources.  For a mean free path which is
shorter than the typical ionizing source separation, large values of
$\langle f_{\rm HI} \rangle_{\rm V}$ can also be produced by the
presence of small, highly ionized regions surrounded by an entirely
neutral IGM (\citealt{Lidz07,AlvarezAbel07}).  We discuss the possible
impact of these pre-ionized regions on near-zone sizes later in this
letter.

The fiducial quasar parameters adopted in the radiative transfer
simulations are an ionizing photon production rate of $\dot N=2\times
10^{57}\rm~s^{-1}$, an age of $t_{\rm Q}=10^{7}\rm~yrs$ and a
power-law spectral index of $\alpha_{\rm s}=1.5$ below the Lyman
limit.  The value of $\dot{N}$ is consistent with the $z>6$ quasar
luminosities inferred by \cite{Fan06t} and independently constrained
by \cite{MesingerHaiman06}.  Quasar ages are rather more
uncertain, although $10^{7}\rm~yrs$ is in agreement within current
observational constraints (\citealt{Martini04}) and theoretical models
(\citealt{Hopkins06b}).  The impact of differing values of $\dot N$
and $t_{\rm Q}$ on the sizes of quasar near-zones is discussed in
detail in BH07a.

\section{Results}
\subsection{A consistent definition for \Lya and \Lyb near-zone sizes} \label{sec:findzone}

There are considerable variations in \Lya and \Lyb near-zone sizes
even for fixed values of $\langle f_{\rm HI} \rangle_{\rm V}$ due to
underlying fluctuations in the IGM density field.  Therefore, a
consistent definition for the sizes of the \Lya and \Lyb near-zones is
very important when comparing individual spectra. \cite{Fan06t}
smoothed the observed \Lya near-zone spectra to a resolution of
$20\rm~\AA$ and identified the near-zone sizes at the first pixel
which dropped below a normalised flux threshold of $F=0.1$.  However,
this definition is not suitable for the \Lyb near-zone, where the
transmission is much weaker and more patchy due to the foreground \Lya
forest.  Alternatively, BH07a measured \Lya and \Lyb near-zone sizes
at the position where the last pixel in an unsmoothed spectrum drops
below a normalised flux threshold of $F=0.1$.  Although providing a
better measure of the \Lyb near-zone size, this definition still
breaks down once the IGM as a whole becomes highly ionized and the
edge of the near-zones become ambiguous.  To overcome these
difficulties, we add the extra condition that the last pixel at which
the flux drops below $F=0.1$ must be followed by a gap of $\Delta z >
0.1$ where the normalised flux remains {\it below} the $F=0.1$
threshold.  Any spectra which do not meet this criterion are
considered to have unidentifiable \Lya and \Lyb near-zones and are
rejected from our analysis.  This occurs for only one and three
sight-lines in the synthetic ESI and HIRES samples, respectively.

\subsection{The impact of ionizing background fluctuations on the
sizes of near-zones}

The sizes of the \Lya and \Lyb near-zones measured from our 480
synthetic ESI spectra are shown as the red diamonds in
Figure~\ref{fig:sizes}.  The top panels show the sizes measured
assuming a uniform ionizing background, while the bottom panels
correspond to the sizes measured with  the assumption of a spatially
fluctuating ionizing background.  We find a substantial scatter in the
near-zone sizes even for fixed assumptions for $\langle f_{\rm HI}
\rangle_{\rm V}$.  The scatter is rather similar for the spectra
computed with either a uniform or spatially fluctuating ionizing
background, suggesting that differences in the density distribution
and radiative transfer effects play a more important role in
generating this variation.  There is also little evidence that
enhanced ionization in the overdense regions increases the average
near-zone size.  Note, however, that during the overlap-phase
spatial fluctuations in the ionizing background are expected to be
larger than we have assumed here and can extend over wider scales.
This is due to the rapid increase of the mean free path for ionizing
photons in the short phase when the ionized regions percolate.

\begin{table}
\centering
 
  \caption{The \Lya and \Lyb near-zone sizes around the five quasars
  with published spectra at $z>6.1$. The sizes have been measured from
  the spectra using the definition given in
  Section~\ref{sec:findzone}.}

  \begin{tabular}{c|c|c|c} \hline Quasar  & $R_{\alpha}$ [proper
  Mpc]  &$R_{\beta}$  [proper Mpc]  & $R_{\beta} $/$R_{\alpha}$   \\
  \hline	
  $\rm J1509-1749$  & $6.4\pm1.2$  & $12.7\pm1.2$ & $2.0$\\
  $\rm J1250+3130$  & $13.6\pm1.2$ & $9.3\pm1.2$  & $0.7$\\
  $\rm J1623+3192$  & $2.6\pm1.0$  & $3.1\pm1.0$  & $1.2$\\
  $\rm J1030+0524$  & $4.8\pm1.2$  & $6.0\pm1.2$  & $1.2$\\
  $\rm J1148+5251$  & $6.2\pm1.4$  & $5.6\pm1.4$  & $0.9$\\
 
 \hline	
\label{tab:sizes}
\end{tabular}

\end{table}

\begin{figure}
	
  \begin{center}
    \psfig{figure=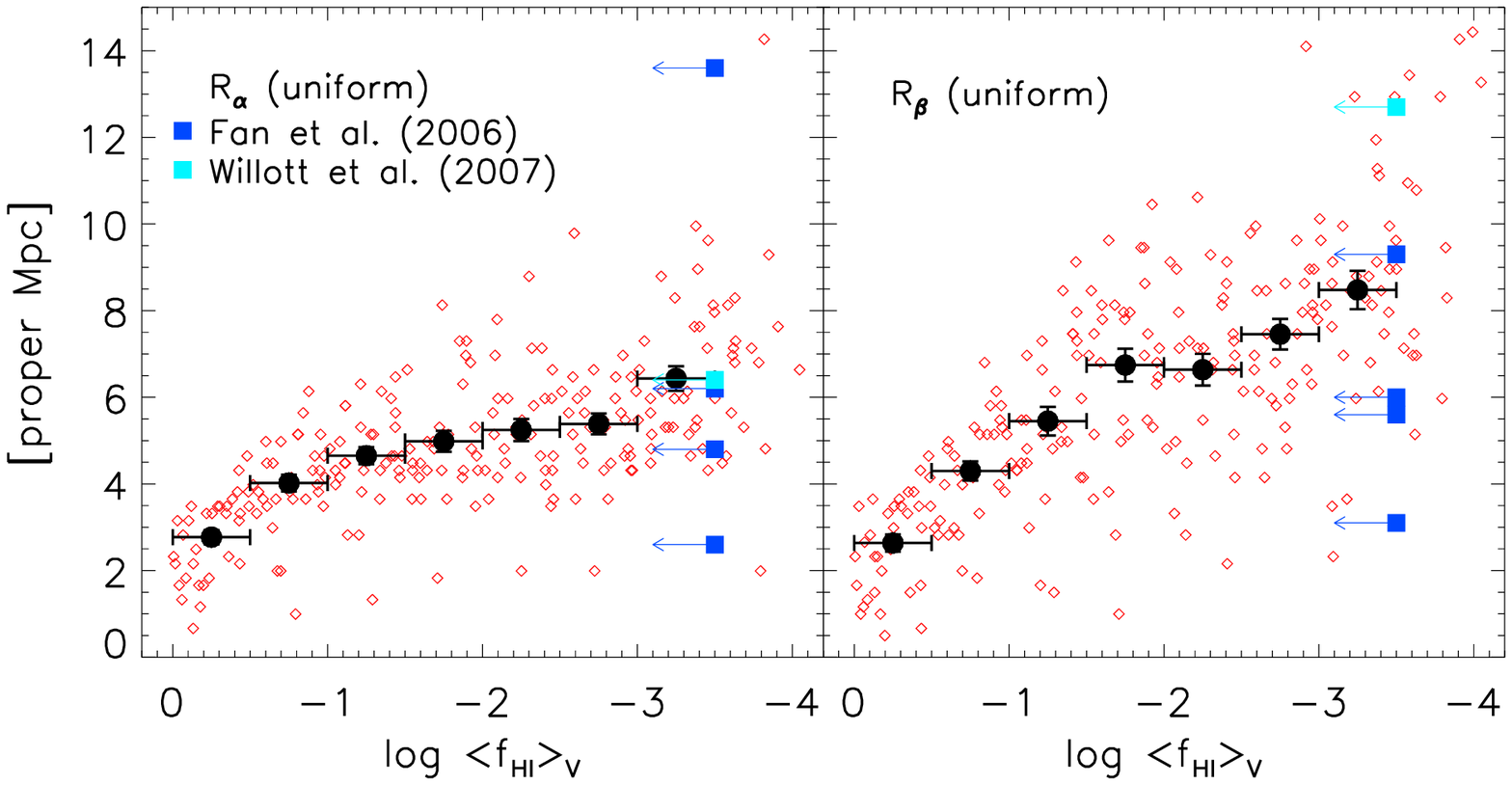,width=0.46\textwidth}
 	\vspace{-5mm}
    \psfig{figure=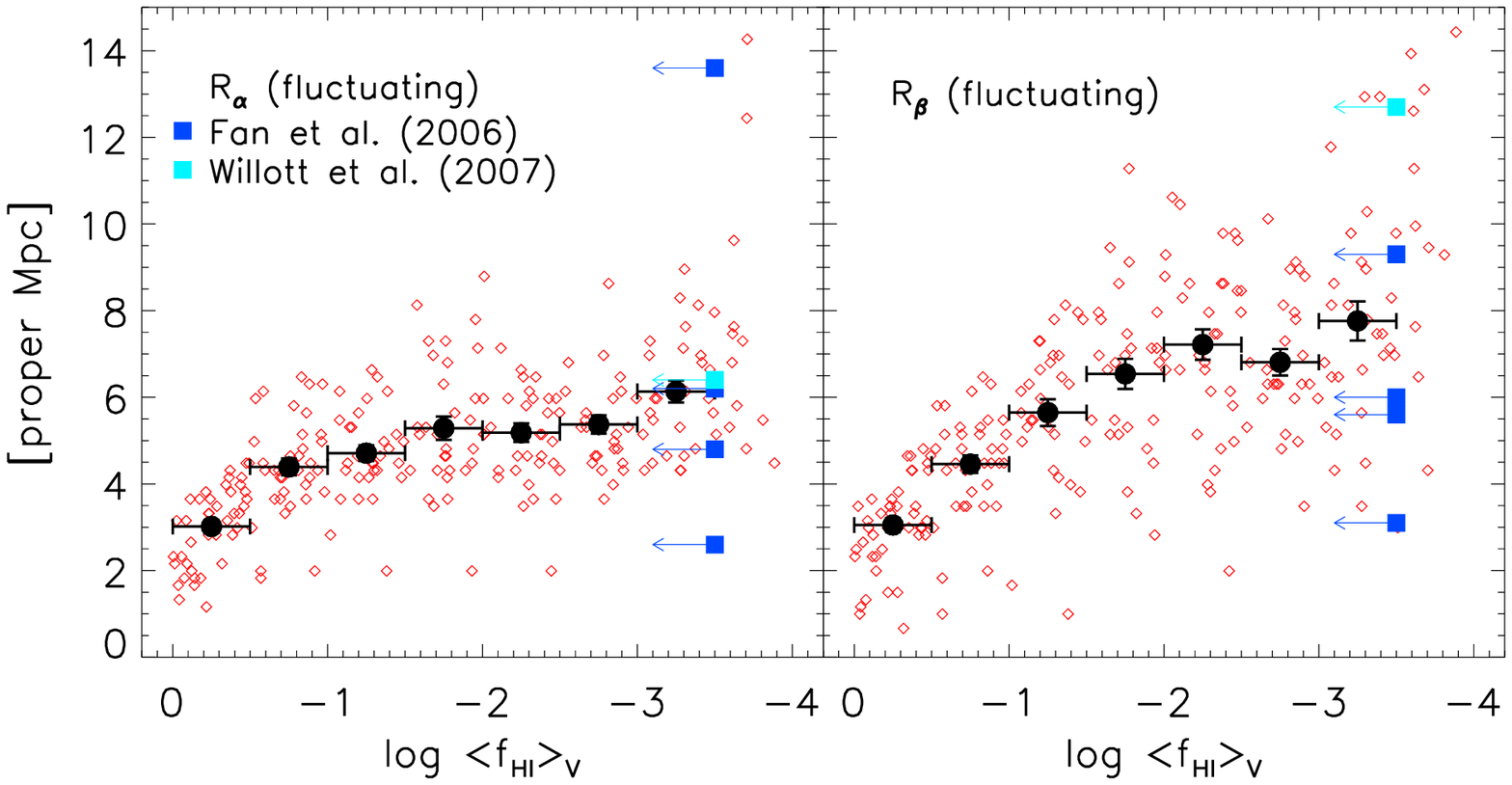,width=0.46\textwidth}
  \end{center}
 
\vspace{5mm} 
\caption{{\it Top:} The red diamonds correspond to \Lya (left panel)
and \Lyb (right panel) near-zone sizes measured from $240$ synthetic
ESI spectra using the method described in Section~\ref{sec:findzone}.
The data are plotted as a function of the volume weighed neutral
hydrogen fraction in the IGM.  A spatially uniform ionizing background
has been assumed.  The filled black circles with $1\sigma$ error bars
correspond to the average near-zone sizes in bins of width
$\Delta[\log \langle f_{\rm HI} \rangle_{\rm V}]=0.5$. {\it Bottom:}
As for the top panel except assuming a spatially fluctuating ionizing
background. The filled blue and cyan squares in all panels correspond
to near-zone sizes we derive from the published observational data,
given in Table~\ref{tab:sizes}. }
\label{fig:sizes}

\end{figure}

The filled black circles with error bars in Figure~\ref{fig:sizes}
correspond to the average near-zone sizes in bins of width
$\Delta[\log \langle f_{\rm HI} \rangle_{\rm V}]=0.5$.  The error bars
correspond to the $1\sigma$ standard error of the mean.  There are
three main regimes in the \Lya near-zone size as a function of  $\log
\langle f_{\rm HI} \rangle_{\rm V}$ for the quasar age and luminosity
adopted here. The first is when  $R_{\alpha} \propto \langle f_{\rm
HI} \rangle_{\rm V}^{-1/3}$ for $\langle f_{\rm HI} \rangle_{\rm
V}>0.1$.  In this regime the \Lya near-zone edge closely corresponds to the
\HII IF.  The second regime is when the near-zone corresponds to a
classical proximity zone and the size is independent of the neutral
fraction; in this case the \HII IF actually lies ahead of the \Lya
near-zone edge but the residual neutral hydrogen behind it is large
enough to produce saturated \Lya absorption.  Lastly, for $\langle
f_{\rm HI} \rangle_{\rm V}<10^{-3.5}$ the measured \Lya near-zone
sizes increase again due to additional transmission originating from
regions now highly ionized by the ionizing background.  The \Lyb
near-zones exhibit a similar trend to the \Lya data, except due to the
smaller \Lyb absorption cross-section they are able to trace the
position of the \HII IF to lower IGM neutral hydrogen fractions,
reaching $R_{\beta} \propto \langle f_{\rm HI} \rangle_{\rm V}^{0}$ at
$\langle f_{\rm HI} \rangle_{\rm V}\simeq 10^{-2}$.  Therefore, as
noted by BH07a (but see also \citealt{MesingerHaiman04} for a
different interpretation), evidence for $R_{\beta}/R_{\alpha}>1$ may
provide an interesting constraint on $\langle f_{\rm HI} \rangle_{\rm
V}$.

Finally, we plot the \Lya and \Lyb near-zone sizes we measured from
the five quasars at $z>6.1$ with published spectra as filled blue and
cyan squares in Figure~\ref{fig:sizes} (\citealt{Fan06t,Willott07}).
We have omitted $\rm J1030+0524$ at $z=6.2$ from this sample since
this is a broad absorption line quasar which complicates the
measurement of the near-zone size.  Following \cite{Fan06t}, the sizes
are rescaled to a common absolute magnitude of $M_{1450}=-27$ by
assuming the near-zone size is proportional to $\dot N^{1/3}$,
although $\dot N^{1/2}$ may be more appropriate (BH07a).  This
magnitude corresponds to $\dot N \simeq 1.9\times 10^{57}\rm~s^{-1}$
for the quasar spectrum adopted in this work.  The lower limit of
$\langle f_{\rm HI} \rangle_{\rm V} \ga 10^{-3.5}$ measured from the
Gunn-Peterson trough limits is assumed.  An uncertainty of $\Delta z =
0.02$ in the systemic redshift of the quasar is adopted when
determining the systematic error on the near-zone sizes.  We do not
show the systematic uncertainties on the measured near-zone sizes in
Figure~\ref{fig:sizes} for clarity, but the data are given in
Table~\ref{tab:sizes}.  The observational data exhibit a large scatter
similar to that seen in the synthetic data, and are consistent with a
wide range of neutral fractions.

\begin{figure}
	
 \begin{center}
    \psfig{figure=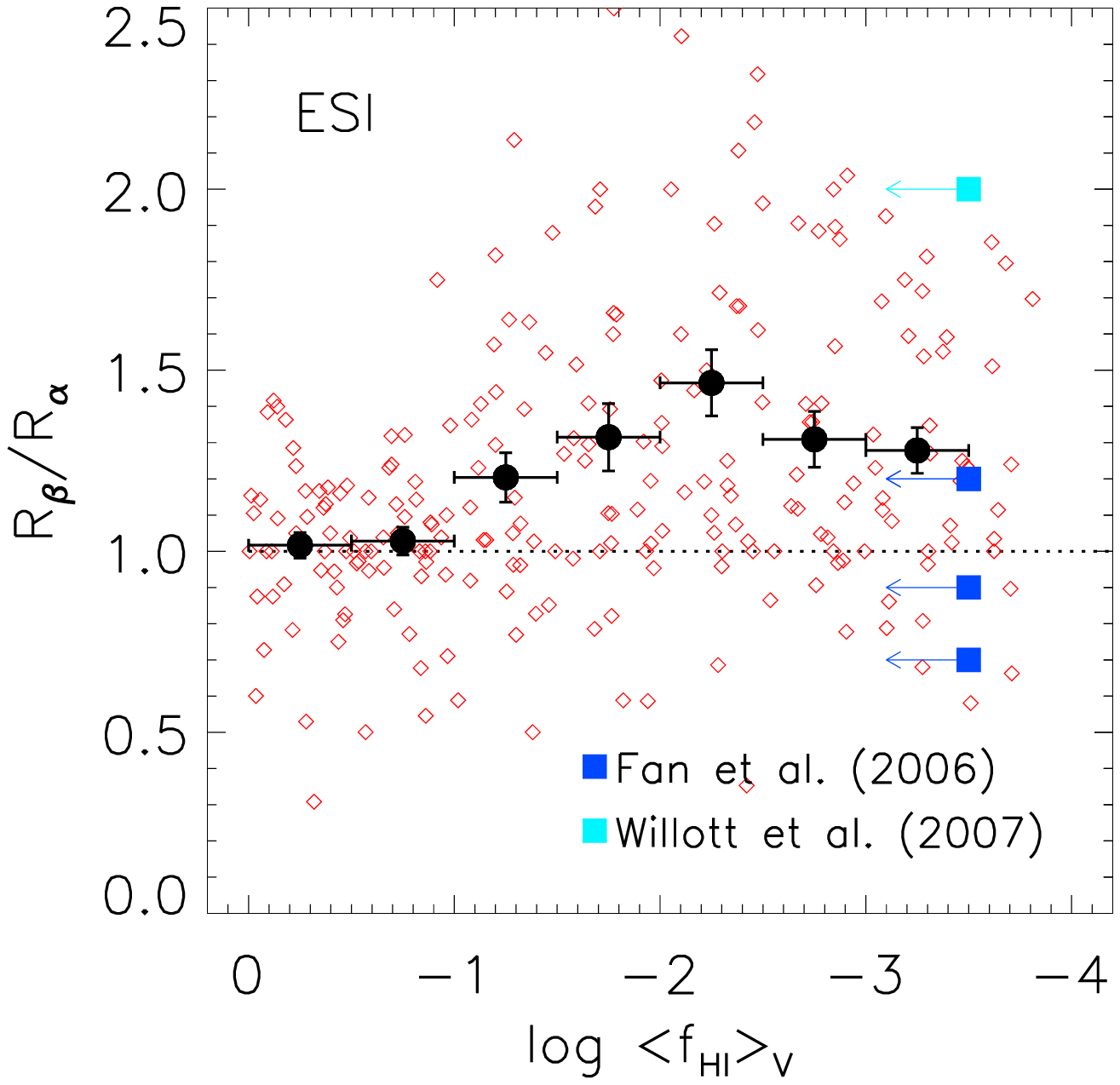,width=0.275\textwidth}
 	\hspace{-15mm}
    \psfig{figure=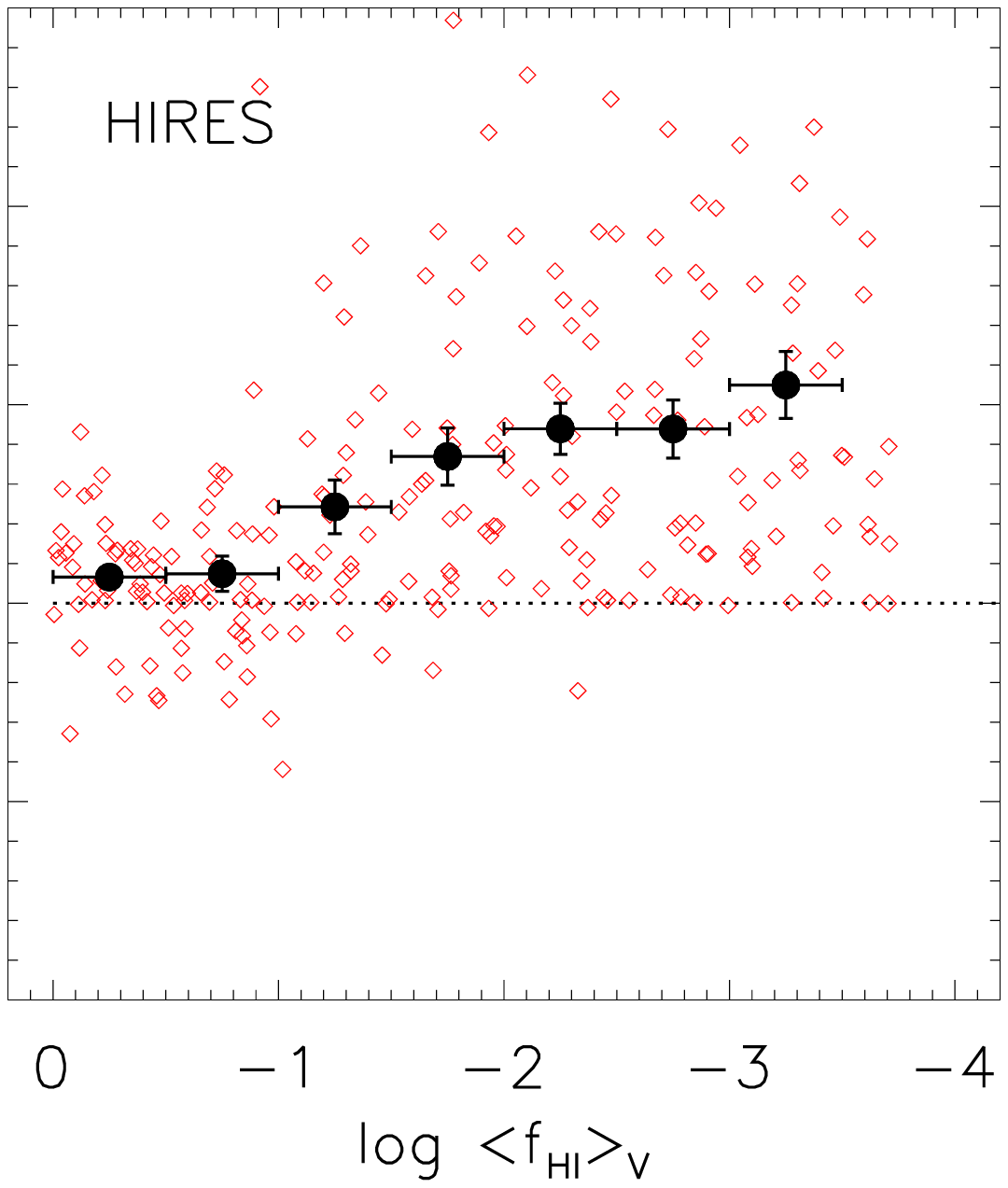,width=0.275\textwidth}
  \end{center}
 
\vspace{-0.5cm} 
\caption{{\it Left:} The ratio of \Lyb to \Lya near-zone sizes as a
function of the volume weighted neutral hydrogen fraction, measured
from synthetic ESI spectra constructed assuming a spatially
fluctuating ionizing background.  The red diamonds show the individual
measurements and the filled black circles with $1\sigma$ error bars
correspond to the averages in bins of width $\Delta[\log \langle
f_{\rm HI} \rangle_{\rm V}]=0.5$.  The filled blue and cyan squares
correspond to the near-zone sizes measured from the published data,
given in Table~\ref{tab:sizes}. {\it Right:} As for the left panel, except
the data are now measured from synthetic HIRES spectra.}
\label{fig:ESI}
\end{figure} 

\subsection{Probing the neutral fraction with the ratio of \Lyb and
\Lya near-zone sizes}

A potentially interesting constraint on $\langle f_{\rm HI}
\rangle_{\rm V}$ may be obtained by considering the ratio of the \Lyb
to \Lya near-zone sizes.  The red diamonds in the left panel of
Figure~\ref{fig:ESI} correspond to $R_{\beta}/R_{\alpha}$ as a
function of  $\log \langle f_{\rm HI} \rangle_{\rm V}$ for the synthetic
ESI spectra constructed using the fluctuating ionizing background
model.  The filled black circles show the average values of
$R_{\beta}/R_{\alpha}$ in bins of width $\Delta[\log \langle f_{\rm
HI} \rangle_{\rm V}]=0.5$.  The error bars correspond to the
$1\sigma$ standard error of the mean.  The earlier analysis by BH07a found
$R_{\beta}/R_{\alpha}>1$ was consistent with $\langle f_{\rm HI}
\rangle_{\rm V} \la 10^{-2}$.  However, this was based on a small
sample of $20$ synthetic spectra which used only four different
initial values of  $\langle f_{\rm HI} \rangle_{\rm V}$.  The improved
statistics and wider sampling of $\langle f_{\rm HI} \rangle_{\rm V}$
in this work indicate $R_{\beta}/R_{\alpha}>1$ when $\langle f_{\rm HI}
\rangle_{\rm V} \la 0.1$ at the $3\sigma$ level for a sample of
around $30$ ESI spectra.

Assuming that $R_{\beta}/R_{\alpha}>1$ is indeed consistent with
$\langle f_{\rm HI} \rangle_{\rm V} \la 0.1$, how reliable a
constraint on $\langle f_{\rm HI} \rangle_{\rm V}$ can the current
observational data provide?  The observed near-zone sizes we measure
from the published data (\citealt{Fan06t,Willott07}) at $z>6.1$ are
plotted as the filled blue and cyan squares in Figure~\ref{fig:ESI}.  Once
again, the observational data are consistent with a wide range of
$\langle f_{\rm HI} \rangle_{\rm V}$.  The mean of the observational
data favours a value of $R_{\beta}/R_{\alpha}= 1.2$, although
with only five data points this result is certainly not a significant
one.  The simulated data indicates that about $20$ ESI spectra would
be required for a $2\sigma$ result.   Obtaining reliable constraints
on $\langle f_{\rm HI} \rangle_{\rm V}$ using this technique with the
current data is probably not yet possible.

Higher resolution quasar spectra would enable a more accurate
determination of  the sizes of the \Lya and \Lyb near-zones.  This is
demonstrated in the right panel of Figure~\ref{fig:ESI} in which  our
simulated spectra  have been processed to resemble data taken with
HIRES.  The scatter in the data is significantly reduced with the
higher resolution spectra.  Strong evidence for a neutral fraction
$\langle f_{\rm HI} \rangle_{\rm V}<10^{-1}$ should be obtainable with
a sample of 30 high-resolution spectra at the $3-5\sigma$ level.

\subsection{The effect of quasar age, ionizing luminosity and surrounding galaxies}

An important caveat to the argument presented in this work is that the
$\langle f_{\rm HI} \rangle_{\rm V}$ threshold where
$R_{\beta}/R_{\alpha}>1$ becomes statistically significant is also
degenerate with the age of a quasar, its ionizing luminosity and the
size of any pre-existing ionized region around the quasar host halo.
The observable $R_{\beta}/R_{\alpha}>1$ corresponds to the $\langle
f_{\rm HI} \rangle_{\rm V}$ threshold at which $R_{\rm
HII}>R_{\alpha}$, where $R_{\rm HII}$ is the extent of the \HII IF.
The position of the \HII IF will scale approximately as $R_{\rm HII}
\propto \langle f_{\rm HI}\rangle_{\rm V}^{-1/3}\dot{N}^{1/3}t_{\rm
Q}^{1/3}$ but the saturated \Lya near-zone size will scale as
$R_{\alpha} \propto \dot{N}^{1/2}$ only (BH07a).  Therefore, if the
mean quasar age is ten times larger than the fiducial $10^{7}\rm~yrs$,
the \HII IF will be around twice as far from the quasar but the \Lya
near-zone size will remain unchanged.  The $\langle f_{\rm HI}
\rangle_{\rm V}$ threshold corresponding to $R_{\beta}/R_{\alpha}>1$
will thus be ten times larger.  Alternatively, a smaller quasar age
will lower the $\langle f_{\rm HI} \rangle_{\rm V}$ threshold
corresponding to $R_{\beta}/R_{\alpha}>1$. Disentangling this quasar
age and neutral fraction degeneracy from the data is potentially very
difficult without a good independent estimate of the typical quasar
age.  Note, however, the degeneracy with $\dot{N}$ is less dramatic,
since both the near-zone size and the \HII IF position scale similarly
with $\dot{N}$. Finally, luminous quasars are expected to be hosted by
rather massive dark matter haloes and the reionization process may be
further advanced in their immediate environment compared to a typical
region of the Universe.  If pre-existing ionized regions produced by
clustered galaxies lower the local value of $\langle f_{\rm HI}
\rangle_{\rm V}$ around these quasars, the IFs will travel further
than one expects for a uniformly ionized IGM
(\citealt{Lidz07,AlvarezAbel07}).  The observed near-zone sizes may
then exhibit $R_{\beta}/R_{\alpha}>1$ even if $\langle f_{\rm HI}
\rangle_{\rm V}>0.1$.

\section{Conclusions}

We have carefully examined the potential of the ratio of \Lyb to \Lya
near-zone sizes as a probe of the IGM \HI fraction at $z>6$ with a
large set of detailed radiative transfer simulations.  Adopting a
robust, simultaneous definition for \Lya and \Lyb near-zone sizes, we
find that a sample of around $30$ ESI quasar spectra would be required
to distinguish between an IGM which has an \HI fraction greater or
less than $10$ per cent at the $3\sigma$ level.  Unfortunately, the
current observational data, consisting of five quasar spectra at
$z>6.1$, are too few to place significant constraints on $\langle
f_{\rm HI} \rangle_{\rm V}$.  Our results also suggest that, at least
in the post-overlap phase of reionization, fluctuations in the
ionizing background will have a negligible effect on near-zone sizes.
However, if the typical quasar age is substantially longer than the
canonical $10^{7}\rm~yrs$, or if pre-existing ionized  regions
surround quasar host haloes during the pre-overlap stage of
reionization, it may still be very difficult to distinguish between a
highly ionized or substantially neutral IGM.  Attempts to disentangle
the former effect from the data may benefit from alternative
statistics, such as the detection of a Gunn-Peterson trough damping
wing (\citealt{MesingerHaiman04,MesingerHaiman06}), while
distinguishing the latter effect may be possible with a sample of
$z>6$ gamma ray burst (GRB) spectra, which should probe less biased
regions compared to luminous quasars.  However, only one such spectrum
has been obtained so far (\citealt{Totani06}) and its interpretation
is hampered by damped \Lya absorption from neutral hydrogen in the GRB
host galaxy.

High-resolution data significantly improves the ability to constrain
the neutral fraction, and HIRES spectra of $z>6$ quasars are already
in existence (\citealt{Becker05}).  Next generation infra-red surveys
(see \citealt{Lawrence07} for a recent review) such as the UKIRT
Infrared Deep Sky Survey (UKIDSS) are expected to find several tens of
new quasars at $z\simeq 6$ (\citealt{Venemans07}).  Future data sets
should thus be able to provide an interesting limit on the IGM neutral
hydrogen fraction from the relative sizes of \Lya and \Lyb near-zones.

\end{document}